\documentclass[conference]{IEEEtran}
\IEEEoverridecommandlockouts
\usepackage{cite}
\usepackage{amsmath,amssymb,amsfonts}
\usepackage{algorithmic}
\usepackage{graphicx}
\usepackage{textcomp}
\usepackage{xcolor}
\usepackage{subfiles}
\usepackage{multirow}
\usepackage{tabularx}
\usepackage{hyperref}
\usepackage{booktabs}
\usepackage{array}
\usepackage{amssymb}

\newcommand{\mypara}[1]{\vspace{1mm}\noindent\textbf{#1}}

\def\BibTeX{{\rm B\kern-.05em{\sc i\kern-.025em b}\kern-.08em
    T\kern-.1667em\lower.7ex\hbox{E}\kern-.125emX}}
\begin{document}

\title{Using LLMs in Software Requirements Specifications: An Empirical Evaluation}

\author{\IEEEauthorblockN{Madhava Krishna\IEEEauthorrefmark{1}$^\dagger$, Bhagesh Gaur\IEEEauthorrefmark{1}$^\dagger$, Arsh Verma$^{\dagger\ddagger}$, Pankaj Jalote$^\dagger$}
\IEEEauthorblockA{$^\dagger$\textit{Department of Computer Science, IIIT Delhi} 
\quad\quad $^\ddagger$\textit{Wadhwani AI}\\
\{madhava20217, bhagesh20558, arsh17221, jalote\}@iiitd.ac.in}
\thanks{$^*$ Equal Contribution}
}

\maketitle

\begin{abstract}
The creation of a Software Requirements Specification (SRS) document is important for any software development project. Given the recent prowess of Large Language Models (LLMs) in answering natural language queries and generating sophisticated textual outputs, our study explores their capability to produce accurate, coherent, and structured drafts of these documents to accelerate the software development lifecycle. We assess the performance of GPT-4 and CodeLlama in drafting an SRS for a university club management system and compare it against human benchmarks using eight distinct criteria. Our results suggest that LLMs can match the output quality of an entry-level software engineer to generate an SRS, delivering complete and consistent drafts. We also evaluate the capabilities of LLMs to identify and rectify problems in a given requirements document. Our experiments indicate that GPT-4 is capable of identifying issues and giving constructive feedback for rectifying them, while CodeLlama's results for validation were not as encouraging. We repeated the generation exercise for four distinct use cases to study the time saved by employing LLMs for SRS generation. The experiment demonstrates that LLMs may facilitate a significant reduction in development time for entry-level software engineers. Hence, we conclude that the LLMs can be gainfully used by software engineers to increase productivity by saving time and effort in generating, validating and rectifying software requirements.
\end{abstract}
\section{Introduction}
One of the fundamental requirements for developing software is clearly defined objectives and communication of the same to the software developers. It is standard practice in the industry to prepare a Software Requirements Specification (SRS) document at the beginning of the project to ensure clear communication between all the stakeholders involved. Traditionally, this document is prepared and reviewed manually with contributions from multiple team members. This practice can take weeks or months but is earmarked as an essential time investment. However, with the advent of Large Language Models (LLMs), we find ourselves at a revolutionary moment where such ML models can assist in automating even the toughest technical tasks. In this work, we explore the efficacy of GPT-4 (ChatGPT) \cite{chatgpt2023}, and CodeLlama \cite{roziere2023code} in generating, validating, and rectifying these documents using natural language prompts describing the features of our intended final product. We also present  evidence on the time saved by utilizing these models for the task.

Software Requirements Specifications are formally defined as a “specification for a particular software product, program, or set of programs that perform certain functions in a specific environment” containing details about the functionality, external interfaces, performance, attributes, and design constraints imposed on an implementation \cite{ieee830-1998, ieee29148-2018}. The SRS is a comprehensive document that serves as the foundational blueprint of the entire software development lifecycle. It ensures alignment among all stakeholders by delineating a software venture's expectations, operational methods, and inherent constraints, thus ensuring clarity regarding the software's intended purpose, functionality, and limitations. This can expedite the development trajectory by preemptively diagnosing and fixing potential pitfalls, obviating the need for extensive rework or modifications. A robust SRS serves as a unified reference that ensures that all parties have synchronized expectations and responsibilities, promoting efficient cooperation.

With the explosion in the adoption of Artificial Intelligence (AI), it is now possible to leverage AI agents in key requirement engineering (RE) tasks like extraction, classification, prioritization, and validation of requirements \cite{yang2022survey, kici2021bert}. The popularization of Large Language Models (LLMs) based on the transformer \cite{vaswani2017attention} architecture, such as Generative Pre-trained Transformer (GPT) \cite{radford2018improving, radford2019language} and the recently released LLaMa \cite{touvron2023llama} models has led to an increase in their usage in RE tasks.

In this paper, we evaluate the proficiency of LLMs like GPT-4 \cite{achiam2023gpt} (ChatGPT) \cite{chatgpt2023} and CodeLlama \cite{roziere2023code} in formulating the SRS for a software project. To demonstrate this, we chose the task of designing a university’s student club management web portal. We grade LLM-generated documents and the human benchmark on various metrics relevant to this task chosen from the literature. Our goal is to evaluate if LLMs can reduce human effort by independently developing comprehensive documents or providing worthy drafts that humans can refine easily. We attempt to identify their weaknesses and strengths in such complex design tasks through an empirical analysis. We also observe their utility for validating and correcting software requirements and analyze the time saved by utilizing them in designing requirements specifications.

Specifically, our study addresses the following research questions (RQ):

\mypara{RQ1:} How do LLMs, with focus on GPT-4 and CodeLlama, perform relative to an entry-level software engineer in SRS creation?

\mypara{RQ2:} How do GPT-4 and CodeLlama perform in validating the quality of requirements and suggesting improvements?

\mypara{RQ3:} What is the reduction in effort by utilizing LLMs for generating SRS documents?

The final prompts, settings, and chats we used for our experiments can be accessed from the following GitHub\footnote[1]{\href{https://github.com/madhava20217/LLMs-for-SRS-Prompts}{ https://github.com/madhava20217/LLMs-for-SRS-Prompts}} repository.
\section{Related Works}

The creation and evaluation of SRS documents have been studied extensively. The importance of having complete, unambiguous, and contextually rich SRS documents has been emphasized repeatedly \cite{clancy1995chaos, kamata2007does, knauss2009investigating}. To facilitate writing good SRS documents, previous work \cite{wiegers1999writing} provides guidelines to write good requirements and proposes measures to evaluate the same. Researchers have also proposed a framework \cite{davis1993identifying} to quantify key aspects of a good SRS and facilitate easier comparison through a weighted score, which our evaluation framework heavily derives from.

Despite taking measures to write good requirements, creating and evaluating SRS documents has inherent challenges rooted in subjectivity, such as ambiguity, inaccuracy, and inconsistency \cite{wilson1999writing}. A taxonomy of commonly observed defects in SRS documents has been prepared \cite{alshazly2014detecting} along with a process to detect such defects by isolating each section of the SRS and searching for typical errors. Further, researchers have studied guidelines to determine the quality of an SRS and created an automated testing tool \cite{fabbrini2000quality} to determine the level of ambiguity of each sentence. 

Prior works have also explored the automated generation of SRS documents. \cite{georgiades2005requirements, georgiades2010automatic} discuss methods to generate SRS documents in natural language using Natural Language Syntax and Semantics Requirements Engineering (NLSSRE) methodology that aims to do requirements discovery, analysis, and requirements specification. \cite{mandal2023large} utilize a BERT \cite{devlin2018bert} model as an encoder and an LSTM \cite{hochreiter1997long} network as a decoder to generate requirements.

In order to of examine and correct an SRS to satisfy critical qualities of correctness, completeness, and unambiguity, among others, \cite{castaneda2012improving, siegemund2012measure, wei2023automatic} use knowledge graph and ontology tracking-based approaches to re-format requirements such that they are consistent, correct, traceable, unambiguous, and organized.

The mainstream adoption of LLMs, which serve as powerful information retrieval and zero-shot generation tools \cite{sun2022investigating, daun2023chatgpt} has led to a transition in the software development lifecycle \cite{pothukuchi2023impact}. It is possible to prompt LLMs in several ways -- zero-shot prompting, prompt chaining, few shot prompting \cite{brown2020language}, chain-of-thought prompting \cite{wei2022chain, kojima2022large}, and tree-of-thought prompting \cite{long2023large, yao2023tree}, among others. In the context of using LLMs for requirements engineering, \cite{arvidsson2023prompt} provided guidelines for prompt engineering, observing the impact of the task context on the quality of documents and that a more detailed context resulted in better outputs for requirements engineering tasks.  \cite{arora2023advancing} experiment with a real-life use case on how LLMs can be used in requirements elicitation, user story generation, quality assurance, and requirement validation. \cite{rahman2024automated} use LLMs, specifically GPT-4 \cite{achiam2023gpt}, to create user stores from an SRS document.
\section{Methodology}

In this section, we describe the process of generating the SRS using LLMs and the instructions given to the human oracle.

\subsection{Task definition}

We chose to generate an SRS to design a university’s student club management portal as the primary task for our experiments. There are four main types of users - the administrator, the student council coordinator, club coordinators, and students. We provide a summary of the key functionalities below:-

The administrator serves as the central authority of the system with complete control and visibility of activities assigned to all system users. They can create, edit, and delete clubs and appoint the student council coordinator and club coordinators from within the registered students. The administrator should be a part of the college staff, being from the Students Affairs department.

The student council coordinator manages and approves events for which each club puts forward proposals. The club coordinators for each club can add or edit the club's information and schedule events and club activities, which will then be approved first by the student council coordinator and then the administrator. They can also manage the club members and the visibility and access for club events to the student body. All other university students should be able to view the schedule of the club events and their details and register for the same. All users should be able to login using the Google sign-in API.

\subsection{Benchmark for SRS Generation}

\begin{figure}[ht]
\centering
\includegraphics[width=0.6\linewidth]{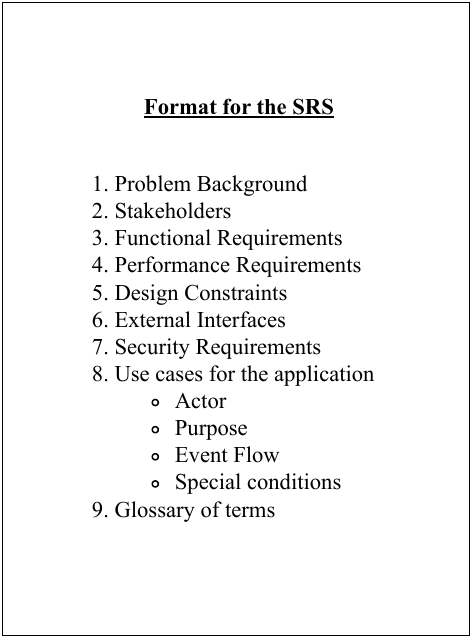}
\caption{
The format of the SRS used for the study.
}
\vspace{-4mm}
\label{fig:srs_template}
\end{figure}

We established a benchmark with a human-generated SRS document, which conformed to IEEE specifications \cite{ieee830-1998}. This document was created and reviewed by software engineering experts. The template used for the SRS is given in figure \ref{fig:srs_template}. The SRS includes the problem background, information about stakeholders, functional and non-functional requirements, and use cases.

We experimented with both iterative prompting in a conversational format and a single, comprehensive prompt. Ultimately, we decided to go with the latter, whilst including a detailed context for all SRS generations since we observed the best results using this. A maximum of two additional prompts were allowed to fix minor errors and formatting.

\subsection{Document generation with GPT-4}
We used OpenAI’s ChatGPT \cite{chatgpt2023}, powered by the GPT-4 \cite{achiam2023gpt} model (25th September 2023 version) to generate the SRS. We explicitly provided both prompt and context to the large language model after observing a noticeable improvement in the detail of the generated documents compared to only prompt or context tuning, aligning with the findings of Arora et al. (2023) \cite{arora2023advancing}. The context contained guidelines for creating good SRS documents. The prompt contained the output format comprising major headings and subheadings as a template and concrete information about the use case, and, apart from minor changes, was maintained constant between ChatGPT and CodeLlama34b (ref. \ref{sec:CodeLlama_generation}).

The context was provided in response to two questions, added in a recent update to ChatGPT:
\begin{enumerate}
    \item \textbf{What would you like ChatGPT to know about you to provide better responses?}
    
    We provided detailed information in the form of guidelines for creating quality requirements documents, following the methodology of \cite{ieee830-1998} and \cite{wiegers1999writing}.
    \item \textbf{How would you like ChatGPT to respond?}
    
    We explicitly prompted the model to be detailed and thorough in its response.
\end{enumerate}

\subsection{Document Generation with CodeLlama}
\label{sec:CodeLlama_generation}

We chose CodeLlama-34b to represent open-source LLMs as an alternative to GPT-4 because of its robust performance. We could not use a larger model due to GPU memory constraints. We obtained the model from the official HuggingFace repository source and used it with the Text Generation WebUI \cite{textgenwebui} to facilitate a ChatGPT-like experience. The model was processed to the GGML \cite{ggml} format and loaded using llama.cpp \cite{llama.cpp}  to allow for faster generation on the CPU. We used a context length of 16,384 with the simple-1 preset for the hyperparameters and set num\_new\_tokens to 4096.

No quantization was used as it was found to induce hallucinations and disorganized responses. We also experimented with smaller models but noted the performance of the 34b model to be the most detailed.

As with ChatGPT, we provided characteristics of a good SRS in the context. As CodeLlama34b does not have limits to the context length, we were able to include more details about each trait. The prompt, however, remained the same.

\subsection{Evaluation Strategy for SRS documents}

To facilitate a strong and unbiased evaluation of the SRS documents, they were anonymized and shared with independent reviewers who were not involved in the generation process. These reviewers rated the documents on various parameters (ref. Table \ref{tab:eval_criteria_srs}). Our recruits included four experts from academia and the industry with at least three years of experience and familiarity with the practices followed in software development.

\begin{table}[t]
{%
\small
\centering
\begin{tabular}{c p{0.8\columnwidth}}
\toprule
Rating & \multicolumn{1}{c}{Interpretation}                                                                      \\ \midrule
1      & Strongly Disagree: Falls far below the expected standards for the particular parameter being evaluated. \\
2      & Disagree: Requires significant improvement to meet the expected standards.                              \\
3 & Neutral: Meets the expected standards for the particular parameter being evaluated, but the document misses some details. \\
4      & Agree: Generally meets or slightly exceeds the expected standards with minor areas for improvement.     \\
5 & Strongly Agree: Excellent and fully meets or exceeds the expected standards for the parameter being evaluated.            \\ \bottomrule
\end{tabular}%
}
\vspace{1mm}
\caption{The grading scale used for evaluating the SRS along with their interpretations.}
\label{tab:rating_interpretations}
\vspace{-7mm}
\end{table}

To evaluate the SRS documents and answer \textbf{RQ1}, we chose parameters prevalent in literature \cite{davis1993identifying, wiegers1999writing}, evaluating the documents on a per-requirement and document-wide basis. The ratings were given on a 5-point Likert scale. The interpretation of each grade is given in Table \ref{tab:rating_interpretations}. Reviewers first graded each SRS using the document-wide metrics shown in Table \ref{tab:eval_criteria_srs}. Next, the reviewers graded each requirement of an SRS in the following sections:  functional requirements, performance requirements, design constraints, external interfaces, and security requirements, based on the metrics in Table \ref{tab:eval_criteria_srs}. We then report the average scores for each section.

All SRS documents were standardized to have the same formatting to reduce human bias during evaluation.

\begin{table}[t]
\tabcolsep=0.08cm
\small
\centering
\begin{tabular}{>{\centering\arraybackslash}m{11mm} p{0.25\columnwidth} p{0.6\columnwidth}}
\toprule
Grading Scheme &
  Parameter &
  Definition \\ \midrule
\multirow{13}{*}{\rotatebox[origin=c]{90}{Per-requirement}} &
  Unambiguous &
  A requirement is unambiguous if and only if it has only one possible interpretation. \\
 &
  Understandable &
  A requirement is understandable if all classes of SRS readers can easily comprehend its meaning with a minimum of explanation. \\
 &
  Correctness &
  A requirement is deemed correct when it accurately represents a required feature or function the system must possess. \\
 &
  Verifiable &
  A requirement is verifiable if finite, cost-effective techniques exist for verifying that it is satisfied by the system as built. \\ \midrule
\multirow{12}{*}{\rotatebox[origin=c]{90}{Document-wide}} &
  Internal Consistency &
  An SRS is internally consistent if and only if no subsets of individual requirements conflict. \\
 &
  Non-redundancy &
  An SRS is not redundant if no requirement is restated more than once. \\
 &
  Completeness &
  An SRS is complete if it details all functions, describes all responses, provides organizational clarity, and avoids placeholder text. \\
 &
  Conciseness &
  An SRS is concise when it delivers all necessary information briefly without sacrificing its quality. \\ \bottomrule
\end{tabular}
\vspace{1mm}
\caption{Evaluation criteria for the SRS. Each section of the SRS was evaluated on the per-requirement measures. The SRS on the whole was evaluated on the document-wide measures.}
\label{tab:eval_criteria_srs}
\vspace{-7mm}
\end{table}

\subsection{Validation and Correction of Requirements}
For the experiments on validating and correcting requirements, that answer \textbf{RQ2}, we prompted the LLMs to validate the quality of each requirement in the human SRS created previously and correct them in the same conversation. We used GPT-4 with the January 2024 version of ChatGPT for this task, and we used the same configuration for CodeLlama34b as we did to generate the SRS. As before, ChatGPT was given explicit instructions to generate a detailed and thorough response.

As with SRS generation, we set both the prompt and context for each LLM. The context contained information about the four parameters for evaluating the requirements: unambiguity, understandability, correctness, and verifiability, as given in Table \ref{tab:eval_criteria_srs} and the format of the SRS (Figure \ref{fig:srs_template}). The prompt contained the SRS without the section on use cases, and the goal was to individually assess requirements on a scale of 1 to 5 as per the descriptions in Table \ref{tab:rating_interpretations} for each parameter provided in the context. Apart from minor changes in the prompts and context to improve the generation, the inputs were the same to both LLMs.

The conversation flow for validating and correcting requirements was as follows:
\begin{enumerate}
    \item Evaluate individual requirements in a section on a scale of 1 to 5 with justifications for the corresponding rating on each parameter.
    \item Reformat the evaluations into a table with the section, the requirement number, the ratings for the four parameters, and the justifications for the ratings.
    \item Prompt the LLM to correct the requirements that did not obtain perfect ratings in any of the four parameters.
\end{enumerate}
\section{Quality of Generated SRS Documents}

\begin{figure}[t]
\centering
\includegraphics[width=\linewidth]{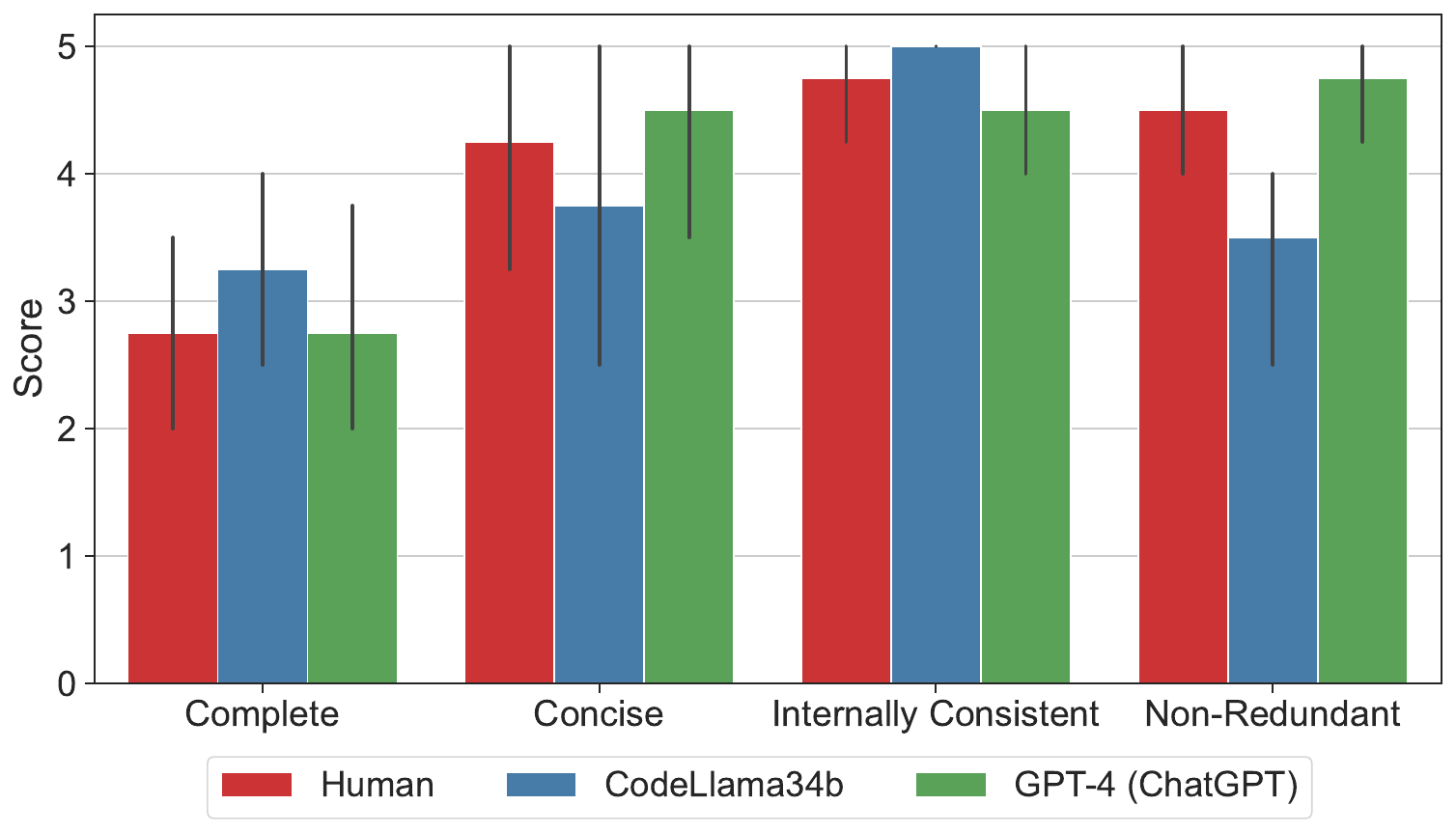}
\vspace{-6mm}
\caption{Overall SRS evaluation. The graph corresponding to document-wide evaluation parameters and has been obtained by averaging the ratings provided by human graders.}
\vspace{-4mm}
\label{fig:srs_overall_results}
\end{figure}

Table \ref{tab:generated_srs_macro_stats} shows a high-level comparison of the three SRS documents, highlighting the length and the number of requirements in each section. We note that CodeLlama generated a shorter document than the human benchmark despite having more requirements than the human benchmark in four out of seven cases. From Fig. \ref{fig:srs_overall_results}, we observe that while ChatGPT generated the most concise and non-redundant document with the least number of requirements in all sections except Functional Requirements, it was not rated to be a complete document.

\begin{table}[t]
\tabcolsep=0.07cm
\small
\centering
\begin{tabular}{@{}clccc@{}}
\toprule
\multicolumn{1}{c}{Area}      & \multicolumn{1}{c}{Section} & Human   & CodeLlama & GPT-4\\ \midrule
\multirow{7}{*}{\rotatebox[origin=c]{90}{SRS-Sections}} & Functional Requirements     & 10      & 8             & 9               \\
 & Performance Requirements & 2  & 8 & 2 \\
 & Design Constraints       & 4  & 8 & 2 \\
 & External Interfaces      & 3  & 5 & 2 \\
 & Security Requirements    & 4  & 7 & 3 \\
 & Use cases                & 10 & 7 & 4 \\
 & Glossary                 & 14 & 6 & 3 \\ \midrule
\multicolumn{2}{c}{Length}                                  & 7 pages & 6 pages       & 3 pages         \\ \bottomrule
\end{tabular}
\vspace{1mm}
\caption{Number of requirements in each section of the SRS documents generated.}
\vspace{-7mm}
\label{tab:generated_srs_macro_stats}
\end{table}

Documents generated by CodeLlama34b were often verbose, detailed, and covered a lot of aspects critical to the software.  On the other hand, ChatGPT generated short, crisp documents that often lacked the detail that the former offered. This is reflected in the completeness, conciseness, and non-redundancy scores in Figure \ref{fig:srs_overall_results}: CodeLlama34b scores the highest in completeness, indicating that it covered the most requirements for the use case. ChatGPT, on the other hand, scores the maximum in conciseness and non-redundancy while performing equal to the human SRS in completeness. We also observe that CodeLlama34b scores the most in internal consistency, indicating minimal conflicts in the requirements it generates. ChatGPT trails the human SRS in this regard.

CodeLlama always formatted requirements in the language as specified in \cite{wiegers1999writing}, while ChatGPT did that only for a few requirements. However, both LLMs mixed up the requirements corresponding to each section despite indicating what they should contain.

\subsection{Per-Requirement Evaluation}

\begin{figure*}
\centering
\includegraphics[width=\textwidth]{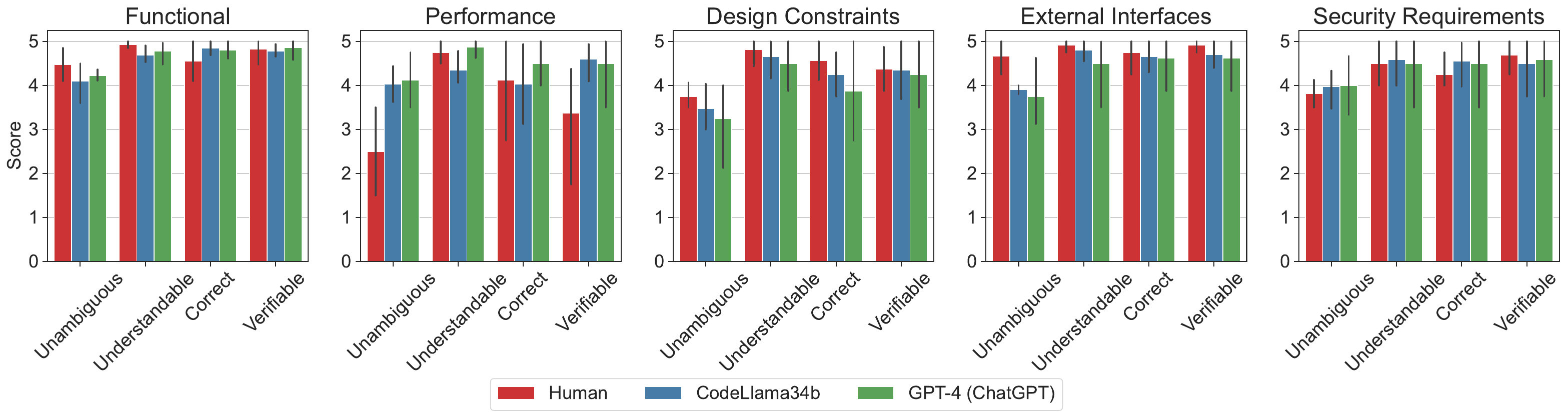}
\caption{Per-requirement evaluation results for the three SRS documents. Each graph corresponds to each section of the SRS and has been obtained by averaging the ratings provided by human graders for that part of the SRS.}
\vspace{-3mm}
\label{fig:srs_per_requirement_results}
\end{figure*}
The summary of the per-requirement grades in Fig. \ref{fig:srs_per_requirement_results} shows that while the first draft of the human SRS is the best document overall, the CodeLlama SRS often comes close and occasionally outperforms it in some categories. The document generated by ChatGPT almost always performs worse than the document generated by CodeLlama. In cases where it does better, CodeLlama never falls too far behind.

\mypara{Functional Requirements:}
All SRS documents were close in all parameters. The human SRS outscored the LLM-generated documents in unambiguity and understandability, but trailed them in terms of correctness. Between the LLMs, we note that ChatGPT performs at par or marginally better than CodeLlama34b. The requirements generated by ChatGPT were atomic, clear, and relevant to the use case. CodeLlama34b, on the other hand, clubbed major requirements into a single requirement despite clear instructions, having a minor impact on the understandability and potentially introducing some ambiguity. However, as the high scores for all documents suggest, the difference was not significant.

\mypara{Performance Requirements:}
Both LLMs demonstrate very strong performance, with ChatGPT taking the lead. Despite ChatGPT delivering fewer requirements than CodeLlama34b, the individual requirements were precise and without ambiguity. CodeLlama34b’s responses, however, catered more to the use case at hand, while ChatGPT and the human SRS contained generic requirements like ``the database should contain K concurrent users". 

\mypara{Design Constraints:}
There is a clear ranking achieved on all parameters, with the human SRS superseding both the LLMs followed by the CodeLlama SRS in second. We observe that CodeLlama34b recycled requirements from other sections, such as security requirements, and often contained irrelevant requirements. ChatGPT failed to be descriptive and provided generic responses.

\mypara{External Interfaces}
Both LLMs made mistakes here and performed much worse than the human benchmark: ChatGPT included the database schema, while CodeLlama mislabeled some design constraints as external interface requirements. Ambiguities were present in the responses, as indicated in the results. However, CodeLlama performs marginally better than ChatGPT.

\mypara{Security Requirements}
ChatGPT generated generic responses, while CodeLlama generated detailed responses. We notice strong performance by LLMs in unambiguity, understandability, and correctness, with CodeLlama leading the rest in 2 of the four parameters. On average, the LLMs performed better than the human benchmark.
\section{Validation and Correction of Software Requirements}

\subsection{Validation of Requirements}

To determine whether LLMs can determine the quality of a requirement, we compare the mean deviations of the ratings provided by LLMs to the averaged ratings from the human graders for each requirement.

    $$deviation = Rating_{LLM} - \bar{Rating}_{Human}$$

\begin{figure*}
\includegraphics[width=\textwidth]{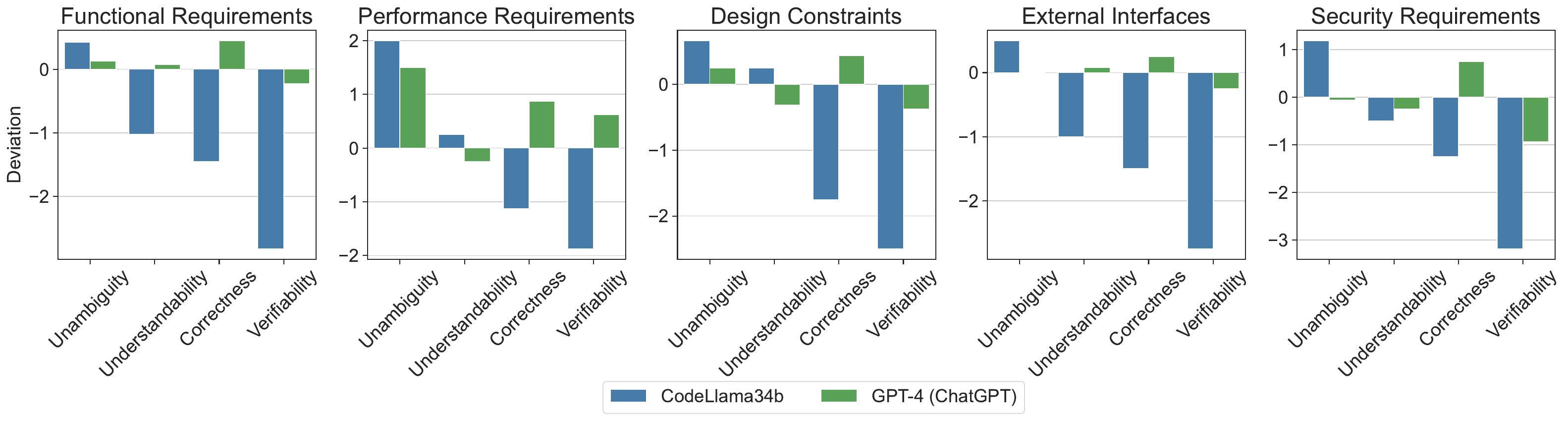}
\vspace{-6mm}
\caption{The mean deviations of the LLM-obtained ratings from human ratings averaged for each section.}
\label{fig:verification_mean_avg}
\end{figure*}

Where $Rating_{LLM}$ is the rating provided by the LLM, and $\bar{Rating}_{Human}$ is the average of the ratings provided by human evaluators for that requirement. A positive deviation suggests that the LLM graded the SRS more optimistically than human graders, and a negative deviation indicates a more pessimistic evaluation. The mean deviations for each section are shown in figure \ref{fig:verification_mean_avg}.

During the generation, CodeLlama omitted one requirement in the design constraints section and two requirements in the external interfaces section from a total of four and three requirements, respectively. The omitted requirements were excluded from the calculation of the mean deviations. No such behavior was observed for ChatGPT.

We observe that ChatGPT consistently has a lower mean deviation than CodeLlama, remaining under $\pm1$ for the vast majority of cases, indicating that it generally agrees with the human grades. CodeLlama gave scores of 4 and 5 to all requirements in unambiguity, significantly poor scores compared to both ChatGPT and the humans for correctness and verifiability, leading to large negative deviations. In addition, the justifications provided were often generic and imprecise, for ex. “Login is a common functionality, therefore it is easy to understand and has only one interpretation, making it unambiguous”, or “This requirement is general and has only one interpretation, making it unambiguous”.

In contrast to CodeLlama, ChatGPT provided verbose and constructive feedback for each parameter, often aligning with the feedback given by human graders. It identified flaws within the performance requirements, a section widely noted by human graders as ambiguous and not verifiable, while also noticing ambiguities in the design constraints and security requirements sections. In addition, it gave high scores to the external interfaces section, which was reported to have the least defects from prior evaluations. It also provided relevant remarks for each evaluation, like “'Considerable amounts' is vague” and “Specific traffic metrics needed” for performance requirements, “'Easy' and 'readable' are subjective” for design constraints, and “Broad, more specifics of attacks needed” for security requirements.

\subsection{Correcting Requirements}

After verification, we prompted the LLMs to rectify any and all requirements that did not receive a perfect score in any of our evaluation parameters. ChatGPT corrected ten out of the 11 requirements it identified. On the other hand, CodeLlama gave imperfect scores to all requirements but corrected just four of them. Tables \ref{tab:correction_chatgpt}, \ref{tab:correction_codellama34b} provide details of some of the original and corrected requirements for ChatGPT and CodeLlama, respectively.

\begin{table*}[t]
\tabcolsep=0.15cm
\small
\footnotesize
\begin{tabular}{
>{\centering\arraybackslash}m{18mm}
>{\centering\arraybackslash}m{15mm}
>{\centering\arraybackslash}m{12mm}
>{\centering\arraybackslash}m{12mm}
>{\arraybackslash}m{55mm}
>{\arraybackslash}m{55mm}
}
\toprule
Section &
  \centering Requirement Number &
  Human scores &
  GPT-4 scores &
  \centering Old Requirements &
  New Requirements \\ \midrule
Performance Requirements &
  4.1 &
  1.75, 4.75, 4, 4.25 &
  3, 4, 5, 3 &
  Platform should withstand considerable amounts of traffic. &
  The platform must support up to 10,000 concurrent users and maintain functionality during peak usage times, such as event registration openings and result announcements. \\ \midrule
Design Constraints &
  5.1 &
  3.25, 4.75, 3.75, 3.5 &
  3, 4, 5, 3 &
  Develop with accessibility features like screen readers, font adjusters, etc. &
  The platform will include accessibility features such as screen readers compatibility, font size adjusters, high contrast mode, and keyboard navigation support. \\ \midrule
Design Constraints &
  5.3 &
  2.5, 4.75, 4.75, 3.25 &
  3, 4, 5, 3 &
  Easy to maintain with readable and well-documented code. &
  The platform's codebase shall adhere to industry-standard practices for readability, including clear naming conventions, modular design, and comprehensive documentation. \\ \midrule
External Interfaces &
  6.1 &
  4.5, 4, 4.5, 4.75 &
  4, 5, 5, 4 &
  Compatibility with major browsers. &
  The platform must be compatible with the latest two versions of major browsers including Chrome, Firefox, Safari, Edge, and Brave. \\ \midrule
Security Requirements &
  7.4 &
  4, 4.25, 4, 4.5 &
  3, 4, 5, 3 &
  User data management as per institute's policy. &
  User data shall be managed in compliance with the institute's Data Protection Policy, including provisions for data encryption, regular audits, and adherence to privacy regulations. \\ \bottomrule
\end{tabular}
\vspace{1mm}
\caption{Original and corrected requirements by GPT-4 (ChatGPT). The parameters for ratings by humans and ChatGPT are unambiguity, understandability, correctness, and verifiability, respectively.}
\label{tab:correction_chatgpt}
\end{table*}
\begin{table*}[t]
\tabcolsep=0.15cm
\footnotesize
\centering
\begin{tabular}{
>{\centering\arraybackslash}m{18mm}
>{\centering\arraybackslash}m{15mm}
>{\centering\arraybackslash}m{12mm}
>{\centering\arraybackslash}m{12mm}
>{\arraybackslash}m{55mm}
>{\arraybackslash}m{55mm}
}
\toprule
Section &
  \centering Requirement Number &
  Human scores &
  CodeLlama scores &
  \centering Old Requirements &
    New Requirements 
  \\ \midrule
Performance Requirements &
  4.2 &
  3.25, 4.75, 3.75, 3.5 &
  4, 5, 3, 1 &
  The platform should be responsive and have an average response time of 300 milliseconds or less under normal load. &
  The platform should respond within a maximum of 500 ms under normal traffic conditions, while ensuring that it maintains the same level of performance for all users. \\ \midrule
External Interfaces &
  6.2 &
  4.75, 4.75, 4.75, 4.75 &
  Not Found &
  The platform would use Google sign-in API to allow users to log in using their institute domain email address. Upon login, the users would be redirected to Google API’s interface. &
  The platform will enable Google authentication and allow users to access the application directly without being required to go through the Google API's interface. This feature enhances security by eliminating the need for users to log into a third-party service before accessing the institute's student management portal. \\ \midrule
Security Requirements &
  7.2 &
  3, 4.25, 4.5, 4.75 &
  5, 4, 3, 1 &
  The platform should prevent packet interception and tampering with outgoing and incoming requests. &
  All data transmitted between the client and server will be encrypted using SSL/TLS to prevent eavesdropping, man-in-the-middle attacks, and other forms of unauthorized access. The platform will also have a mechanism in place that prevents tampering with outgoing and incoming requests by verifying their authenticity using digital signatures. \\ \bottomrule
\end{tabular}
\vspace{1mm}
\caption{Original and corrected requirements by CodeLlama34b The parameters for ratings by humans and CodeLLaMa34b are unambiguity, understandability, correctness, and verifiability, respectively.}
\vspace{-1mm}
\label{tab:correction_codellama34b}
\end{table*}

Both LLMs’ corrections improve the overall quality of the requirements, making them more comprehensive and reducing ambiguities, as seen from the tables. However, CodeLlama drastically changes performance requirement 4.2, increasing the threshold of response times from the original requirement.
\section{Impact on Effort}

In this section, we answer \textbf{RQ3}. To understand the impact of using LLMs on the effort required for developing an SRS, we performed the generation experiment multiple times for an empirical analysis. 
To determine the time consumed to generate an SRS document with LLMs, we created SRS documents using CodeLlama13b and  ChatGPT with GPT-4 for 4 use cases of various complexities. We recorded the number of iterations and time-per-iteration for context-tuning, prompt-tuning, and prompt-context-tuning to obtain the total effort. We used CodeLlama13b instead of CodeLlama34b as in the other experiments primarily to improve the generation speed. We expect the general results to hold with the larger CodeLlama34b model as well. To determine the human effort required, we engaged three students who generated the SRS documents for these use cases and reported the effort required for each document. The details of the tasks and the average time to set the prompts and contexts are given in Tables \ref{tab:time_taken_chatgpt} and \ref{tab:time_taken_codellama13b} for ChatGPT and CodeLlama13b, respectively.

The following strategy was used for the SRS generation:

\begin{enumerate}
    \item \textbf{Set initial context and prompt.} The context contained information about the task, the expected output, and the level of detail. The prompt had concrete details about the use-case.
    \item \textbf{Context tuning:} Tune the context till the format is correct.
    \item \textbf{Prompt tuning:} Tune the prompt till all details about the task are included.
    \item \textbf{Prompt-context tuning:} Adjust both prompt and context till the SRS contains all required information in the right format.
\end{enumerate}

\begin{table*}[t]
\tabcolsep=0.145cm
\footnotesize
\centering
\begin{tabular}{@{}lcccccccccccc@{}}
\toprule
  \multicolumn{1}{c}{\multirow{2}{*}{Task}} &
  \multirow{2}{*}{Complexity} &
  \multicolumn{2}{c}{Setup time} &
  \multicolumn{3}{c}{Avg. \#Re-iterations} &
  \multicolumn{3}{c}{Avg. Tuning Time} &
  \multirow{2}{*}{Overall time} &
  \multirow{2}{*}{Approx. time for Human SRS} &
  \multirow{2}{*}{Speedup} \\
\multicolumn{1}{c}{}           &        & P    & C    & P    & C    & P + C & P    & C    & P + C &         &      &        \\ \midrule
Club Management Portal & Med & 9.33 & 12   & 1.67    & 2.33    & 0.67  & 4.67 & 3.35 & 1     & 38.67   & 840   & 21.7x \\
Project management portal      & High   & 11.67 & 10.33    & 0.67 & 2 & 0.33  & 2    & 4.33 & 5  & 33.33   & 1440   & 43.2x \\
Golf-scores tracker            & Easy   & 10 & 10.33 & 0.5 & 2.33 & 0.33     & 6    & 2.89    & 4     & 31   & 240    & 7.7x \\
Sports council website         & Med & 9   & 12   & 1    & 2    & 0     & 3    & 4    & 3     & 28      & 720   & 25.7x \\ \midrule

\multicolumn{10}{l}{Average}                                                                                             & 32.75 & 810 & 24.6x \\ \bottomrule
\end{tabular}
\vspace{1mm}
\caption{Time saved by using GPT-4 (ChatGPT) for SRS creation, measured in minutes. \textit{P} represents prompt tuning, \textit{C} represents context tuning, and \textit{P+C} represents both prompt and context tuning.}
\vspace{-7mm}
\label{tab:time_taken_chatgpt}
\end{table*}
\begin{table*}[t]
\tabcolsep=0.145cm
\footnotesize
\centering
\begin{tabular}{@{}lcccccccccccc@{}}
\toprule
  \multicolumn{1}{c}{\multirow{2}{*}{Task}} &
  \multirow{2}{*}{Complexity} &
  \multicolumn{2}{c}{Setup time} &
  \multicolumn{3}{c}{Avg. \#Re-iterations} &
  \multicolumn{3}{c}{Avg. Tuning Time} &
  \multirow{2}{*}{Overall time} &
  \multirow{2}{*}{Approx. time for Human SRS} &
  \multirow{2}{*}{Speedup} \\
\multicolumn{1}{c}{}           &        & P    & C    & P    & C    & P + C & P    & C    & P + C &         &      &        \\ \midrule
Club Management Portal & Med & 6.67 & 16   & 1    & 2    & 0.33  & 1.75 & 2.67 & 2     & 29.67   & 840   & 28.3x \\
Project management portal      & High   & 7.67 & 9    & 0.33 & 1.33 & 1.33  & 5    & 1.75 & 7.25  & 30.33   & 1440   & 47.5x \\
Golf-scores tracker            & Easy   & 5.33 & 9.67 & 0.33 & 0.33 & 0     & 2    & 3    & -     & 16.67   & 240    & 14.4x \\
Sports council website         & Med & 13   & 11   & 0    & 1    & 0     & -    & 4    & -     & 26      & 720   & 27.7x \\ \midrule

\multicolumn{10}{l}{Average}                                                                                             & 25.6675 & 810 & 29.5x \\ \bottomrule
\end{tabular}
\vspace{1mm}
\caption{Time saved by using CodeLlama13b for SRS creation, measured in minutes. \textit{P} represents prompt tuning, \textit{C} represents context tuning, and \textit{P+C} represents both prompt and context tuning.}
\vspace{-3mm}
\label{tab:time_taken_codellama13b}
\end{table*}

Compared to human-created documents, which took between 4 to 24 hours to create once the requirements were specified, LLM-generated documents, while hard to get right in the first go, required considerably less time. The time saving is almost 7-47x from the human-created documents. We also note that more complicated use cases require more time and effort than simpler ones, as reflected in the total time to generate an SRS. However, it should be pointed out that SRS development, in general, consumes only a small portion of the overall software development effort. Hence, the impact of savings in the requirements activity will have only a small impact on the overall project effort and cost.

\section{Conclusion and Discussion}

Software requirements specification (SRS) is a key task in any software project. It is also known to be human-effort-intensive.
In this paper, we report how LLMs can be used effectively to get a strong SRS document --- both through generation and validation ---
and how the use of LLMs can help reduce effort and improve quality.

We note good performance by the LLMs for generating SRS documents (RQ1). The generations are consistent and usually complete. CodeLlama34b tends to create an SRS close to the quality of a document created by an advanced computer science student while often avoiding the inconsistency, ambiguity, and time required for formatting the document into a suitable format. While ChatGPT also performs well, we observe it to generate shorter, less-detailed documents. CodeLlama34b can also create requirements catered to the use case, while ChatGPT's responses were more generic.

For validating and improving requirements (RQ2), we noticed that ChatGPT outperformed CodeLlama34b. It was able to accurately identify more requirements to be corrected while also providing better explanations on the quality. Both of them performed similarly when correcting requirements.

However, getting good results on the first attempt is often hard. The stochasticity of LLMs outputs complicates the generation process and in determining which combination of the prompt and context works successfully. There are a large number of parameters that can be changed to alter the generation, with different configurations resulting in major differences in the output. Additionally, most LLMs are also prone to hallucinations in their responses. During our testing, we observed that while CodeLlama34b generated more verbose and detailed outputs, it was more susceptible to such issues. ChatGPT worked in reverse: its responses were more precise and were less prone to hallucinations, but the compactness often led to the responses being incomplete.

Despite these limitations and the effort taken for the initial setup, LLMs present the potential multi-fold time saving compared to a novice software engineer (RQ3). Hence, with increasing experience in using LLMs for software development, we vouch for their use in tasks such as requirements engineering.
\section{Future Work}

Our work primarily focuses on CodeLlama34b and ChatGPT, which were, at the time of the study, pure-text LLMs. A number of new and improved LLMs have been released since then, with text-vision models steadily gaining traction. Benchmarking their performance on SRS creation and requirement validation can be a possible extension of our work. 

Further, we used general, out-of-the-box models that had not been finetuned. The LLMs often provided general suggestions, especially in the case of non-functional requirements. While they still provided competitive results compared to the human baseline, finetuning these LLMs to create a specialist model could improve the generation quality and accuracy and help incorporate domain-specific knowledge information for more competent generations.

Finally, the prospect of SRS validation, which involves checking the document to ensure it satisfies qualities, needs a more in-depth exploration. While we observe that LLMs can be used for this task, research still needs to be done on developing guidelines to prompt LLMs to do this task efficiently and effectively.

\bibliography{main.bib}

\end{document}